\documentclass[aps,pra,twocolumn,amsmath,amssymb,nofootinbib,superscriptaddress]{revtex4-1}

\usepackage{times}
\usepackage[pdftex]{graphicx}
\usepackage{dcolumn}
\usepackage{bm}
\usepackage{amsmath}
\usepackage{indentfirst}
\usepackage{float}
\usepackage[colorlinks]{hyperref}
\usepackage[dvipsnames]{xcolor}
\usepackage{xcolor}
\usepackage{subfigure}
\usepackage{algorithm}
\usepackage{algorithmicx}
\usepackage{algpseudocode}
\usepackage{amsmath}
\usepackage{verbatim}
\usepackage{makecell}
\usepackage[normalem]{ulem}

\newcommand{\loss}{{\rm loss}}

\newcommand{\ppt}{\vert\Upsilon\rangle}

\newcommand{\trace}{{\rm tr}}

\newcommand{\evo}{\mathcal{U}}
\newcommand{\transfer}{T}
\newcommand{\mem}{\mathcal{C}}
\newcommand{\memsize}{\mathcal{D}}

\newcommand{\qmap}{\mathcal{E}}
\newcommand{\prepare}{\mathcal{P}}
\newcommand{\measure}{\mathcal{M}}
\newcommand{\ptensor}{\Upsilon}

\newcommand{\hnu}{Key Laboratory of Low-Dimensional Quantum Structures and Quantum Control of Ministry of Education, Department of Physics and Synergetic Innovation Center for Quantum Effects and Applications, Hunan Normal University, Changsha 410081, China
}

\begin{document}

\title{Reconstructing Non-Markovian Open Quantum Evolution From Multi-time Measurements}

\author{Chu Guo}
\email{guochu604b@gmail.com}
\affiliation{Henan Key Laboratory of Quantum Information and Cryptography, Zhengzhou,
Henan 450000, China}
\affiliation{\hnu}



\begin{abstract}
For a quantum system undergoing non-Markovian open quantum dynamics, we demonstrate a tomography algorithm based on multi-time measurements of the system, which reconstructs a minimal environment coupled to the system, such that the system plus environment undergoes unitary evolution and that the reduced dynamics of the system is identical to the observed dynamics of it. The reconstructed open quantum evolution model can be used to predict any future dynamics of the system when it is further assumed to be time-independent. We define the memory size and memory complexity for the non-Markovian open quantum dynamics which characterize the complexity of the reconstruction.
\end{abstract}

\maketitle

\section{Introduction}

A quantum system is almost inevitably affected by some environment, in which case the dynamics has to be described in the context of an open quantum system~\cite{InesAlonso2017}. A powerful tool to study open quantum system is the Quantum Map (or Quantum Channel), which is a linear and completely positive (CP) mapping from a quantum state at a time $t_0$ to another quantum state at a later time $t_1$, denoted as $\qmap_{1:0}$~\cite{SudarshanRau1961,JordanSudarshan1961}. Given a microscopic description for the unitary evolution of the quantum system plus an environment~\cite{LeggettZwerger1987}, a reduced Quantum Map acting only on the system can be computed by tracing out the environment. However the details of the environment affecting the quantum system may not be clear in priori, which is usually the case for noisy near-term quantum devices~\cite{AruteMartinis2019,WuPan2021,ZhuPan2022}. Nevertheless, given experimental access to prepare arbitrary initial state of the system and measure the system later, the unknown Quantum Map can also be systematically reconstructed using quantum process tomography (QPT)~\cite{ChuangNielsen1997,ArianoPresti2001}.

However, the Quantum Map can not fully characterize the non-Markovian open quantum dynamics, for example if we consider the Quantum Map between $t_0$ and another time $t_2>t_1$ in the non-Markovian case, the equality $\qmap_{2:0} = \qmap_{2:1}\qmap_{1:0}$ does not hold in general~\cite{RivasPlenio2010,HouOh2011}. In other words, to characterize the quantum dynamics between $t_0$ and any time $t$, one may have to perform a QPT separately for each $t$, which is of course undesirable.
In such situations, a natural question to ask is: given preparation and measurement accesses to the underlying quantum system, can we build a model which fully characterizes the non-Markovian open quantum dynamics of it (for example to predict the quantum state at arbitrary times)? 

The first step to answer this question is to give an informationally complete description of the non-Markovian open quantum dynamics. For this purpose, we look at the classical stochastic process as a reference, which describes a sequence of random variables $X_{k:0} = X_kX_{k-1}\cdots X_{0}$ (the starting time in literatures is usually chosen as $-\infty$ since one is often concerned with stationary stochastic process, but here we choose to be $0$ for correspondence with the quantum case)~\cite{Shalizi2001}. A Markovian stochastic process can be fully characterized by the transition matrix: $P(X_k|x_{k-1})$, with $x_{k-1}$ a specific state at time $k-1$.
In comparison, a non-Markovian stochastic process should be characterized by the conditional probabilities on the all the possible histories: $P(X_k|x_{k-1:0})$, where $x_{k-1:0} = \{x_0, \dots, x_{k-1}\}$ denotes a specific history. Given these facts, the connection between a quantum process and a classical stochastic process can be easily drawn as follows. The quantum state $\rho_k$ at time step $k$ is similar to the random variable $X_k$. The Quantum Map $\qmap_{k:k-1}$ is similar to the transition matrix since it is the current state $\rho_k$ conditioned on the last preparation $\prepare_{k-1}$, which can thus be denoted as $\qmap_{k:k-1} = \rho_k(\prepare_{k-1})$. Given these correspondences, it is clear that to fully characterize the non-Markovian quantum dynamics, a mapping $\rho_k(\Lambda_{k-1:0})$ corresponding to $P(X_k|x_{k-1:0})$ is still in need, which is the current state $\rho_k$ conditioned on a sequence of history quantum operations $\Lambda_{k-1:0}=\{\Lambda_{0}, \dots, \Lambda_{k-1}\}$ at $k$ different times $\{t_0, \dots,t_{k-1}\}$. Since each quantum operation $\Lambda_j$ can be implemented by a measurement  $\measure_{j}$ followed by a preparation $\prepare_{j}$~\cite{MilzModi2017}, this map can also be denoted as $\rho_k(\prepare_{k-1:0}, \measure_{k-1:0})$. The last expression actually closely resembles a special instance of classical stochastic process, the \textit{transducer with memory}, which models a system that emits a random variable $Y_j$ (corresponds to $\measure_j$) given an input random variable $X_{j}$ (corresponds to $\prepare_{j}$) at each time $j$ (in the mean time some ``hidden memory state'' changes which corresponds to the collapse of quantum state upon measurement), and is fully characterized by the conditional probability $P(Y_{k}|X_{k-1:0}, Y_{k-1:0})$~\cite{Shalizi2001}. The mapping $\rho_k(\Lambda_{k-1:0})$ is exactly a $k$-step \textit{process tensor} as discovered recently, which is a linear and CP map from a sequence of quantum operations $\Lambda_{k-1:0}$ to the output quantum state $\rho_k$, moreover, the process tensor represents the most generic quantum measurements one could possibly perform on a quantum system~\cite{CostaShrapnel2016,PollockModi2018a}.

The conditional probability $P(X_k | x_{k-1:0})$ and the process tensor $\rho_k(\Lambda_{k-1:0})$ fully characterize a classical stochastic process and a quantum process respectively. However, these descriptions alone are not efficient since the possible histories, namely $x_{k-1:0}$ and $\Lambda_{k-1:0}$,  grow exponentially with $k$. In the classical case, this problem is solved by constructing a \textit{predictive model} from the observed data $P(X_k | x_{k-1:0})$ (ideally using only a finite $k$). The $\epsilon$-machine is an outstanding predictive model~\cite{ShaliziCrutchfield2001}, which also belongs to the broader class of hidden Markov models.
Briefly, instead of storing all the $P(X_k | x_{k-1:0})$ for any $k$, the $\epsilon$-machine divides the histories $x_{k-1:0}$ into disjoint classes, denoted as $\epsilon(x_{k-1:0})$. Each class (referred to as a causal state) represents all the histories that give the same current state, namely $P(X_k|x_{k-1:0}) = P(X_k|x'_{k-1:0})$ for all $x'_{k-1:0}\in \epsilon(x_{k-1:0})$. In the quantum case, a natural predictive model exists, which is referred to as the open quantum evolution (OQE) model and is defined as follows: the system interacts with an (unknown) environment, such that the system plus environment undergoes unitary evolution and that the observed non-Markovian quantum dynamics is the reduced dynamics of the system after tracing out the environment. However, it is currently unknown how to reconstruct an OQE model based on experimentally measurable quantities (the process tensor), such that the non-Markovian quantum dynamics of the system is fully characterized.

This gap is filled in this work. Concretely, we first present an efficient algorithm for process tensor tomography, the complexity of which grows exponentially with the memory complexity (which will be defined later), but only linearly with $k$. Based on the reconstructed process tensor, we then show that the hidden OQE model can be obtained with little overhead.

\begin{figure}
\includegraphics[width=\columnwidth]{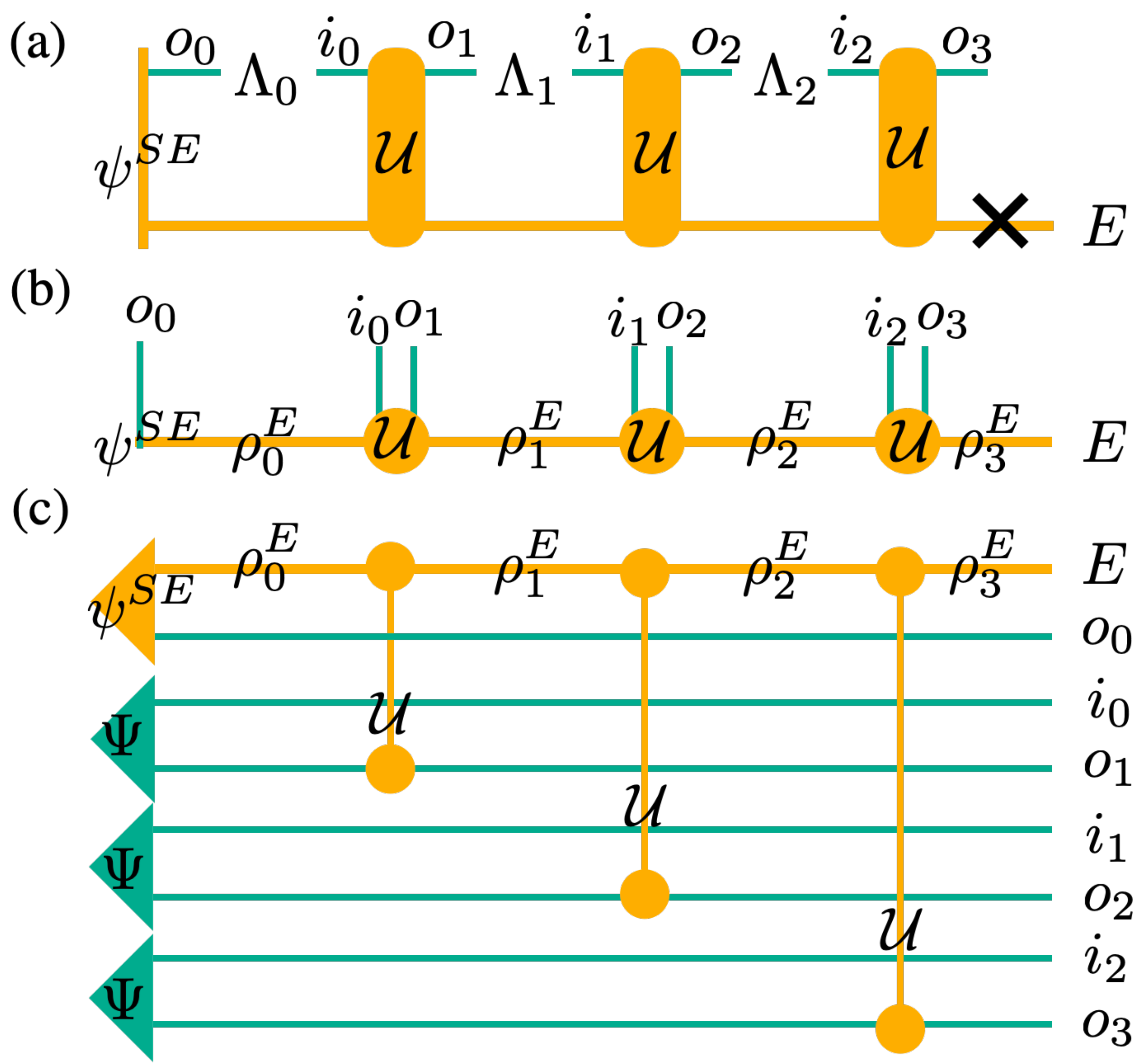}
\caption{(a) Demonstration of a $3$-step process tensor under a hidden OQE model, for which $\evo$ is the unitary evolutionary operator for the system plus the environment, and $\vert \psi^{SE}\rangle$ is the system-environment initial state. $i_j$ and $o_j$ are the $j$-th input and output indices respectively. (b) The Matrix Product State representation of the purified process tensor, for which the environment is not traced out in the end. (c) The quantum circuit implementation of the purified process tensor as a many-body pure state, where $\vert \Psi\rangle$ is the maximally entangled state. The process tensor is obtained by tracing out the environment index. $\rho^E_j$ in (b,c) denotes the $j$-th effective environment state in Eq.(\ref{eq:rho}).
}
\label{fig:fig1}
\end{figure}

\section{The purified process tensor}\label{sec:ppt}
Before presenting the main results of this work, we will first define the purified form of the process tensor (PPT) in the following which will be useful for the later proofs.

A $k$-step process tensor, denoted as $ \ptensor_{k:0} = \rho_k(\Lambda_{k-1:0})$, is implicitly defined based on a hidden OQE model~\cite{PollockModi2018a}:
\begin{align}\label{eq:ptdef}
\rho_k(\Lambda_{k-1:0}) = \trace_E\left( \evo_{k:k-1} \Lambda_{k-1} \dots \evo_{1:0} \Lambda_0 \rho_0^{SE} \right),
\end{align}
where $\rho_0^{SE}$ is the system-environment (SE) initial state and $\evo_{j:j-1}$ is the SE unitary evolutionary operator from time step $j-1$ to $j$, namely $\evo_{j:j-1} \rho^{SE} = U_{j:j-1} \rho^{SE} U_{j:j-1}^{\dagger}$ with $U_{j:j-1}$ a unitary matrix. We further assume $\rho_0^{SE}$ to be a pure state: $\rho_0^{SE} = \vert \psi^{SE}\rangle \langle \psi^{SE}\vert $. This assumption does not loss any generality since if $\rho_0^{SE}$ is a mixed state, one can purify it by adding external degrees of freedom (DOFs) and enlarge $\evo$ accordingly.
The process tensor is demonstrated in Fig.~\ref{fig:fig1}(a). Eq.(\ref{eq:ptdef}) also straightforwardly inspires a tomography algorithm for the process tensor which is similar to the standard QPT: one prepares $\Lambda_{k-1:0}$ with each $\Lambda_j$ selected from an informationally complete set and then perform quantum state tomography on the output quantum state~\cite{WhiteHill2021}. For system size $d$, the total number of configurations grows as $d^{4k+2}$ (since each $\Lambda_j$ lives in a linear space of size $d^2\times d^2$ and $\rho_k$ lives in a space of size $d\times d$), which will soon become unfeasible for relatively small $k$ (Currently the largest experiment for process tensor tomography uses $k\leq 3$~\cite{WhiteModi2020,XiangGuo2021,GoswamiCosta2021}).

The PPT can be defined similar to Eq.(\ref{eq:ptdef}), but without tracing out the environment index in the end, which is demonstrated in Fig.~\ref{fig:fig1}(b).
We can see that the PPT is similar to a pure quantum state and can be naturally written as a Matrix Product State (MPS) 
\begin{align}\label{eq:ppt}
\vert \Upsilon_{k:0}\rangle =& \sum_{o_{k:0}, i_{k-1:0}, \alpha_{k-1:0}}  B^{o_0}_{\alpha_0} B^{i_0,o_1}_{\alpha_0,\alpha_1} \dots B^{i_{k-1},o_k}_{\alpha_{k-1},\alpha_k} \nonumber \\ 
& \times \vert \alpha_k\rangle \vert o_{k:0}, i_{k-1:0}\rangle,
\end{align}
where $o_{k:0} = \{o_0, \dots, o_k\}$ and similarly for $i_{k-1:0}$ and $\alpha_{k-1:0}$. $i_{j}$, $o_j$ are the ``physical indices'' which correspond to particular choices of basis for $\prepare_j$ and $\measure_j$ respectively, $\alpha_j$ is the ``auxiliary index'' corresponding to a choice of basis for the environment after the $j$-th step. The site tensor $B_{\alpha_0}^{o_0} = \langle o_0, \alpha_0 \vert \psi^{SE}\rangle $ and $B_{\alpha_{j-1}, \alpha_j}^{i_{j-1}, i_j}$ with $j\geq 1$ is simply a redefinition of the unitary matrix $U_{j:j-1}$. It is often convenient to view each site tensor of the PPT as a list of matrices $B^{i_{j-1},o_j}$, which are labeled by the physical indices $i_{j-1}, o_j$ and act on the auxiliary index (environment) only ($B^{o_0}$ is a list of vectors). The \textit{bond dimension} of an MPS is defined as the size of each auxiliary index $\alpha_j$, which characterizes the size of each site tensor. The MPS representation in Eq.(\ref{eq:ppt}) is naturally right-canonical since each site tensor satisfies the right-canonical condition~\cite{Orus2014,Schollwock2011}
\begin{align}\label{eq:rightcanonical}
\sum_{i_{j-1}, o_j, \alpha_j} B_{\alpha_{j-1}, \alpha_j}^{i_{j-1}, o_j} (B_{\alpha_{j-1}', \alpha_j}^{i_{j-1}, o_j})^{\ast} = \delta_{\alpha_{j-1}, \alpha_{j-1}'},
\end{align}
where we have used the unitary property of $U_{j:j-1}$. 
The process tensor can be obtained from the PPT by
\begin{align}\label{eq:ppt_to_pt}
\ptensor_{k:0} = \trace_{E}\left(\vert \Upsilon_{k:0}\rangle \langle \Upsilon_{k:0}\vert\right) =\sum_{\alpha_k} \langle\alpha_k\vert \Upsilon_{k:0}\rangle \langle \Upsilon_{k:0}\vert \alpha_k\rangle  .
\end{align}
It has been shown that the process tensor $\ptensor_{k:0}$ for an environment with size $D$ can be written as a Matrix Product Density Operator (MPDO) with bond dimension $D$~\cite{PollockModi2018a}. Moreover, Eq.(\ref{eq:ppt_to_pt}) shows that the process tensor is a very special MPDO, for example given the PPT with bond dimension $D$, one can obtain an MPDO with bond dimension $D$ by Eq.(\ref{eq:ppt_to_pt}), however, an MPDO with bond dimension $D$ in general can not be purified into the form of PPT with the same bond dimension~\cite{VerstraeteCirac2004,CuevasCirac2013,JarkovskyCirac2020,Guo2022b}. This speciality of the process tensor is central to the efficient tomography algorithm we propose for it, which will be shown later.

Interestingly, the PPT can also be implemented using the quantum circuit shown in Fig.~\ref{fig:fig1}(c) (We note that in the original definition of the quantum circuit SWAP gates have been used since physically only the system $o_0$ may directly interact with the environment~\cite{PollockModi2018a}), which converts the PPT defined at multiple times into a multi-qubit many-body quantum state. This can be seen by verifying the outcome of each gate operation simultaneously acting on $o_j$ and the environment index
\begin{align}\label{eq:Tmat}
\evo \vert \Psi\rangle =& \evo\vert \alpha\rangle \frac{1}{\sqrt{d}} \sum_{j=1}^d \vert j \rangle_i \vert j \rangle_o = \frac{1}{\sqrt{d}}\sum_{j=1}^d \vert j\rangle_i (\evo \vert \alpha\rangle\vert j\rangle_o)  \nonumber \\
=& \frac{1}{\sqrt{d}} \sum_{\beta, k, j} U^{j, k}_{\alpha, \beta} \vert \beta\rangle \vert k\rangle_o \vert j\rangle_i ,
\end{align}
which is indeed the $j$-th site tensor of the PPT up to a factor $1/\sqrt{d}$ (Here the state $\vert j\rangle$ with subscript $i$ or $o$ means that it corresponds to the input or output index). Thus the output of the quantum circuit in Fig.~\ref{fig:fig1}(c) without tracing out the environment is exactly the PPT up to a factor $(1/\sqrt{d})^k$.
We also note that the quantum circuit in Fig.~\ref{fig:fig1}(c) is actually a way to prepare a given MPS on a quantum computer, known as the sequentially generated multi-qubit state~\cite{SchonWolf2005}.



\section{Efficient OQE reconstruction algorithm}\label{sec:alg}
In the following we will first present an efficient algorithm to reconstruct the PPT based on experimental measurements, and then we show that the OQE can be reconstructed based on the obtained PPT with little or no additional effort. Since the process tensor is simply related to the PPT (see Eq.(\ref{eq:ppt_to_pt})), it can also be straightforwardly computed based on the obtained PPT.

\begin{figure}
\includegraphics[width=0.9\columnwidth]{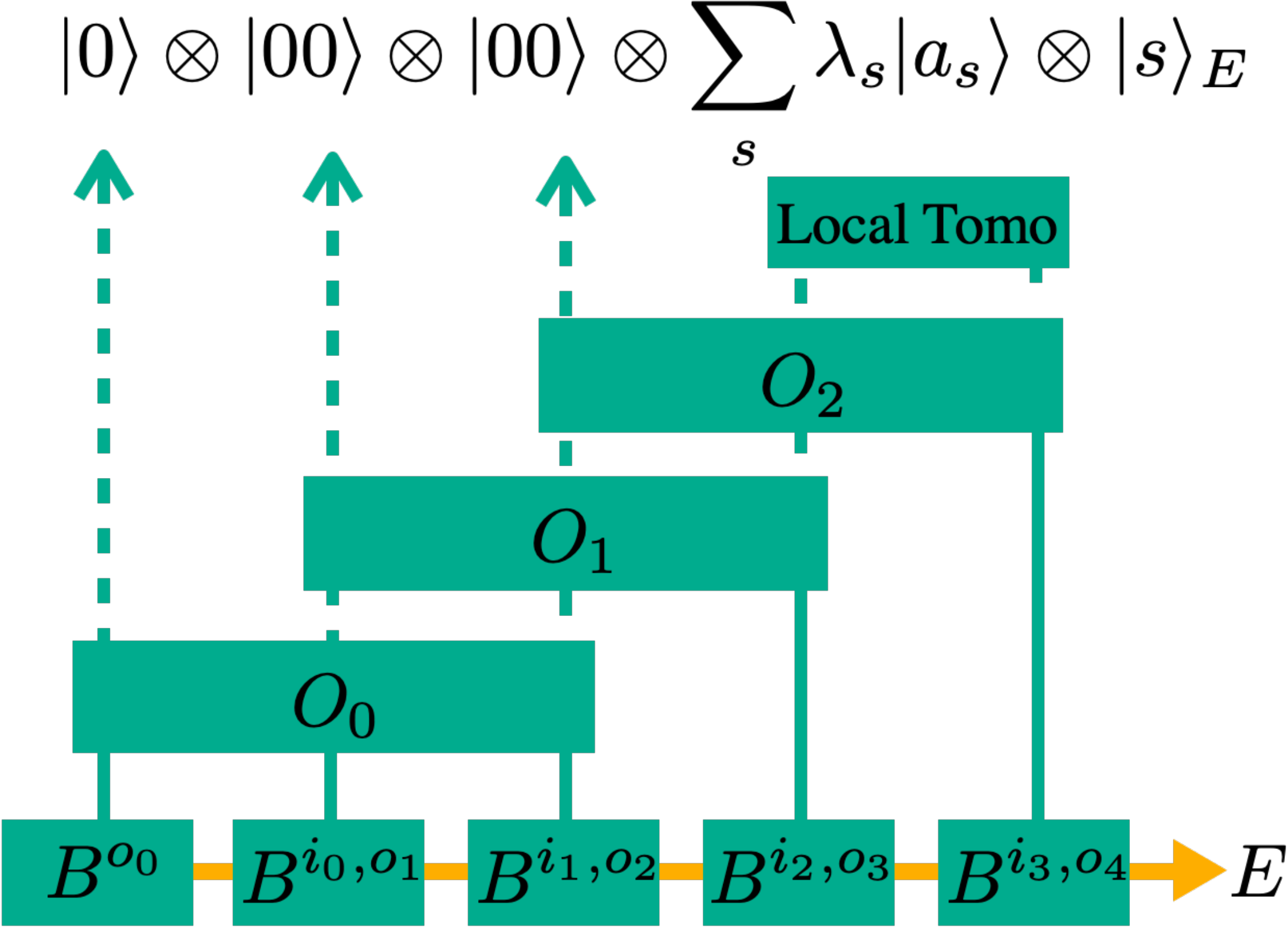}
\caption{Demonstration of the efficient process tensor tomography algorithm for a $4$-step purified process tensor, where we have assumed that the environment size $D\leq d^4$. The unitary gate $O_j$ is the $j$-th disentangling operator in Eq.(\ref{eq:disentangle}). $\vert a_s\rangle$ is the $s$-th eigenstate of the reduced density matrix on the last two sites, and $\lambda_s$ is the square root of the corresponding eigenvalue. 
}
\label{fig:fig2}
\end{figure}

The quantum circuit implementation of the (purified) process tensor as a many-body state immediately motivates to apply the techniques used for many-body quantum state tomography for process tensor tomography. It has been show that for pure or fairly pure (mixed quantum state which can be written as the sum of a few pure states~\cite{GrossEisert2010}) quantum states there exists efficient tomography algorithms with guaranteed convergence, which scales only polynomially with the system size~\cite{CramerLiu2010,BaumgratzPlenio2013,LanyonRoos2017}. These algorithms, however, do not work for a generic MPDO since the latter could easily represent highly mixed quantum states even with a very small bond dimension ($1$ for example). Nevertheless, as we have pointed out, the process tensor is a very special MPDO, and we will show that it allows efficient tomography if one assumes that the size of the unknown environment is bounded by some integer $D$.

Instead of directly reconstructing the process tensor, we will reconstructed the PPT instead, which can be done by tomography of the output quantum state of the quantum circuit in Fig.~\ref{fig:fig1}(c). At first sight this seems unlikely due to the existence of the environment index in the circuit, which is assumed to be not directly accessible. However, as will be shown, we could select a particular environment based only on measurements of the system since it does not affect the process tensor (The PPT itself does not directly correspond to experimentally measurable quantities and could be dependent on the environment basis). 
Following Ref.~\cite{CramerLiu2010}, and assuming that the unknown environment has a size bounded by $D$, one can apply a disentangling quantum circuit onto the (unknown) PPT and get 
\begin{align}\label{eq:tomo}
& O_{f} \dots O_1 O_0 \ppt \vert \psi^{SE}\rangle =  \vert 0\rangle_{f:0} \otimes \sum_{s}\lambda_s \vert a_s\rangle \vert b_s \rangle_E,
\end{align}
with $\kappa = \lceil\log_{d^2}(D)\rceil + 1$, $f = k-\kappa +1$ and $k$ the total number of time steps (sites). $\vert 0\rangle_{f:0}$ means the product state of $\vert 0\rangle$ for site $0$ and $\vert 00\rangle$ for sites $1$ to $f$. $O_j$ is the $j$-th disentangling gate acting on $\kappa$ sites from $j$ to $j+\kappa-1$, defined as~\cite{CramerLiu2010}
\begin{align}\label{eq:disentangle}
O_j = \sum_{r=0}^{d^2-1} \sum_{r'=0}^{d^{2(\kappa-1)}-1} \vert r\rangle_1 \otimes \vert r'\rangle_{\kappa:2} \langle \phi_{rd^{2(\kappa-1)} + r'+1} \vert_{\kappa:1} ,
\end{align}
where $\vert r\rangle_1$ denotes computational basis for site $j$ and $\vert r'\rangle_{\kappa:2}$ the computational basis for sites $j+1$ to $j+\kappa-1$, $\vert \phi\rangle_l$ denotes the $l$-th eigenstate of the reduced density operator $\Upsilon_{j+\kappa-1:j}$ obtained by tomography of sites $j$ to $j+\kappa-1$ after the previous gate operations $O_{0}$ to $O_{j-1}$ have been applied (The $\vert \phi\rangle_l$s are sorted according to the eigenvalues from large to small).
$\vert a_s\rangle$ and $\lambda_s^2$ are the eigenstates and eigenvalues of $\Upsilon_{k:k-\kappa+1}$, which can be obtained by tomography of the last $k-\kappa$ sites after all the gate operations have been applied.
$\vert b_s\rangle_E$ is a set of unknown orthogonal states for the environment. Since we can arbitrarily change the environment basis without any observable effects (it will be traced out when computing the process tensor or any observables), we can simply choose $\vert b_s\rangle_E$ as the computational basis, denoted as $\vert s\rangle_E$. Therefore the quantum state on the right hand side of Eq.(\ref{eq:tomo}) is fixed and we can easily obtain the PPT as an MPS on a classical computer by applying the inverse of the disentangling quantum circuit onto this state. This process tensor tomography algorithm requires $k-\kappa +2$ local quantum state tomography on $\kappa$ sites, as demonstrated in Fig.~\ref{fig:fig2}, the complexity of which is $O((k-\kappa +2) \times d^{2\kappa})$.

From Eq.(\ref{eq:ppt}), the hidden OQE model can be straightforwardly obtained once we have obtained the MPS for a $k$-step PPT. In principle one only needs to prepare the obtained MPS into the right-canonical form and then the site tensors will naturally reveal the hidden OQE model, that is, the first site tensor is the SE initial state and the rest site tensors are the SE unitary evolutionary operators. 
However, in general a site tensor satisfying the right-canonical condition does not guarantee that it is a unitary matrix as in Eq.(\ref{eq:ppt}) up to the factor $1/\sqrt{d}$ (the unitary property of the site tensor implies the right canonical condition in Eq.(\ref{eq:rightcanonical}) but the reverse is not true), the latter is only guaranteed by the physics: the obtained MPS has to result from some hidden OQE model since it is the most general description of an open quantum dynamics~\cite{InesAlonso2017}. In practice, if one losses some precision during the process tensor tomography, this property will not exactly hold. Nevertheless, we could enforce a unitary SE evolutionary operator for each time step by an additional maximally likelihood estimation (MLE). Concretely, one can first find the unitary matrix $\tilde{U}_{j:j-1}$ closest to each $B^{i_{j-1}, o_j}_{\alpha_{j-1}, \alpha_j}$ obtained from the process tensor tomography (this step is not necessary but could be helpful to obtain a good starting point for the next step), then one can further optimize each $\tilde{U}_{j:j-1}$ by minimizing the loss function 
\begin{align}\label{eq:loss}
\loss(\tilde{U}_{1:0}, \dots, \tilde{U}_{k:k-1}) = \left| \vert \tilde{\Upsilon}_{k:0} \rangle - \vert \Upsilon_{k:0} \rangle \right|^2,
\end{align}
where $\vert \cdot \vert^2$ means the square of the Euclidean norm. $\vert \Upsilon_{k:0} \rangle$ means the PPT obtained by process tensor tomography and $\vert \tilde{\Upsilon}_{k:0} \rangle$ is the predicted PPT obtained by substituting all $\tilde{U}_{j:j-1}$ into Eq.(\ref{eq:ppt}). We note that the MLE procedure is purely done on a classical computer. 

\begin{figure}
\includegraphics[width=\columnwidth]{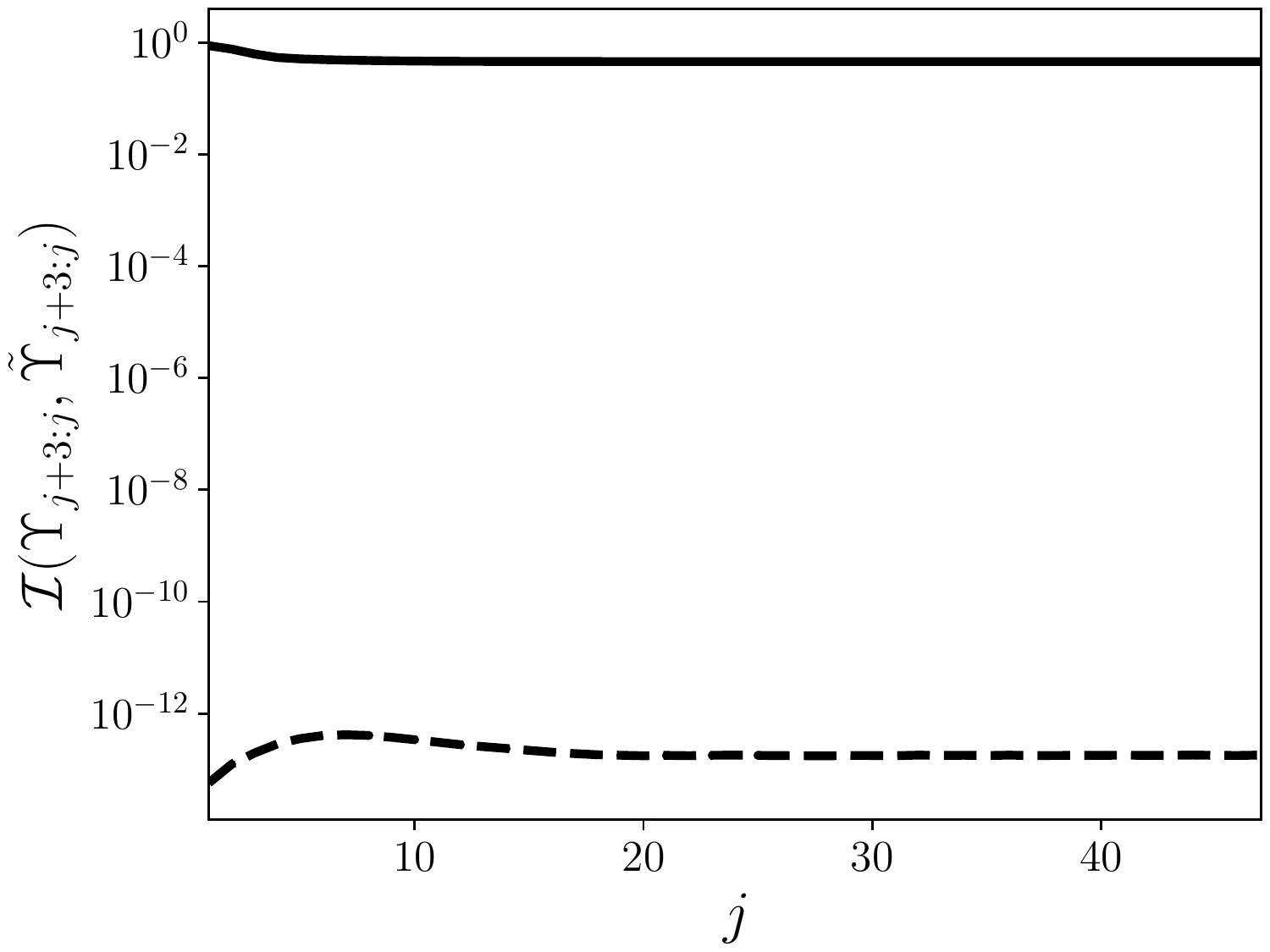}
\caption{Demonstration the tomography algorithm for reconstructing the hidden OQE model. The $x$-axis is the time step $j$. The $y$-axis is the infidelity $\mathcal{I}=1-\mathcal{F}$ ($\mathcal{F}(\rho,\sigma) = \trace^2(\sqrt{\sqrt{\rho}\sigma\sqrt{\rho}})$ is the quantum fidelity between two mixed states $\rho$ and $\sigma$) between the reduced $4$-step process tensor $\Upsilon_{j+3:j}$ computed from the exact OQE model, and $\tilde{\Upsilon}_{j+3:j}$ predicted from the reconstructed OQE model. The black solid and dashed lines correspond to the results obtained by minimizing the loss function in Eq.(\ref{eq:lossc}) with $k=2,3$ respectively. The non-Markovian open quantum dynamics of the system (with $d=2$) is generated by a hidden OQE model (with $D=5$), for which $U = \exp\left(I^{SE} + \eta H^{SE}\right) $ with $\eta=0.1$, $I^{SE}$ the identity matrix and $H^{SE}$ a random Hermitian matrix generated by the normal distribution, and a separable system-environment initial state is used: $\vert \psi^{SE}\rangle = \vert \psi^{S}\rangle\otimes \vert \psi^{E}\rangle$ where $\vert \psi^{S}\rangle$ and $\vert \psi^{E}\rangle$ are randomly generated. 
}
\label{fig:fig3}
\end{figure}

In certain cases one may assume that the environment that is affecting the system does not change with time, which means that there exists an OQE model with a constant evolutionary operator for any time step. Under this assumption, one can again first obtain a $\tilde{U}$ that is closest to some $B^{i_{j-1}, o_j}_{\alpha_{j-1}, \alpha_j}$ with a large $j$, and then obtain the optimal OQE model by minimizing the loss function
\begin{align}\label{eq:lossc}
\loss_c(\vert\tilde{\psi}^{SE}\rangle, \tilde{U}, \tilde{U}^E) = \left| \tilde{U}^E \vert \tilde{\Upsilon}_{k:0} \rangle - \vert \Upsilon_{k:0} \rangle \right|^2,
\end{align}
where $\vert\tilde{\psi}^{SE}\rangle$ is a parameterized pure SE initial state (corresponding to $\tilde{U}_{1:0}$ in Eq.(\ref{eq:loss})) and $\vert \tilde{\Upsilon}_{k:0} \rangle$ is obtained by substituting $\vert\tilde{\psi}^{SE}\rangle$ and $\tilde{U}$ into Eq.(\ref{eq:ppt}), $\tilde{U}^E$ is a parameterized unitary matrix acting on the environment only and is added to compensate the specific choice of basis for the environment made during the process tensor tomography. $\tilde{U}^E$ is not needed in Eq.(\ref{eq:loss}) since it can be absorbed into $\tilde{U}_{k:k-1}$.

The reconstruction algorithm for the OQE is demonstrated in Fig.~\ref{fig:fig3}, from which can see that an equivalent OQE to the original physical one can be learned as long as the time step $k$ used in Eq.(\ref{eq:lossc}) is large enough (we can see that $k=3$ in Eq.(\ref{eq:lossc}) is already enough when we assume the OQE model is time-independent). As a proof of principle demonstration of the reconstruction algorithm we have not considered the noises during the process tensor tomography. We have also used the strategy in Ref.~\cite{ReckBertani1994} to parameterize a general unitary matrix during our numerical simulation.


\section{Memory size and memory complexity}
In the following we will define the memory size and memory complexity for the non-Markovian open quantum dynamics, which are directly related to the complexity of reconstructing the hidden OQE model. These two quantities also characterize the quantum process defined in Eq.(\ref{eq:ptdef}) and are deeply related to the $\epsilon$-machine.

Before introducing these two concepts, we first note that it has been shown that a classical stochastic process can also be simulated by the OQE model on a quantum computer, referred to as the q-simulator~\cite{BinderGu2018}. 
The q-simulator is a more efficient description of the classical stochastic process in that the environment size in a q-simulator could be exponentially smaller than the number of causal states required by an $\epsilon$-machine~\cite{ElliottGu2018,ElliottGu2020}. Moreover, the q-simulator has a one-to-one correspondence with an infinite MPS (iMPS) representation~\cite{YangGu2018} (In the classical case one is often interested in the stationary stochastic process, which would be described by an iMPS, a non-stationary stochastic process will correspond to a finite MPS instead as considered in this work).

Drawing the similarity to the q-simulator and the iMPS representation for the classical stochastic process, we define the \textit{memory size} of a quantum process after time step $j$, denoted as $\memsize_j$, as the Schmidt rank of the PPT in Eq.(\ref{eq:ppt}) at the $j$-th bond (the leg corresponding to $\alpha_j$):
\begin{align}\label{eq:memsize}
\memsize_j = \dim(\alpha_j),
\end{align}
which is the size of the  minimal environment at the $j$-th bond which generates the next evolution.
We also define the \textit{memory complexity} of a quantum process as the Quantum Renyi entropy of the PPT~\cite{YangGu2018}
\begin{align}\label{eq:renyi} 
\mem^{\gamma}_j = \frac{1}{1-\gamma} \log_2\left(\trace\left( \Upsilon_{j:0}^{\gamma}  \right)     \right),
\end{align}
where $\Upsilon_{j:0}= \trace_{E, k:j+1}(\vert \Upsilon_{k:0}\rangle \langle \Upsilon_{k:0}\vert ) = \trace_{k:j+1}(\ptensor_{k:0}) $ is the reduced density matrix for time steps $0$ to $j$ (the first partial trace is taken over the environment index plus the site indices from $j+1$ to $k$).

The memory complexity defined in Eq.(\ref{eq:renyi}) can also be interpreted as the entanglement entropy of an \textit{effective environment state} $\rho^E_j$ after time step $j$, which carries all the history information before (and include) the $j$-th time step. Concretely, $\rho^E_j$ is defined recursively as
\begin{align}\label{eq:rho}
\rho^E_j = \overleftarrow{\transfer_j}(\rho^E_{j-1}) = \sum_{i_{j-1}, o_j} \left(B^{i_{j-1}, o_j}\right)^{\dagger} \rho^E_{j-1} B^{i_{j-1}, o_j},
\end{align}
with $\rho^E_0 = \trace_S(\rho^{SE}_0)$ and $\transfer_j$ the $j$-th transfer matrix of the PPT: $\transfer_j = \sum_{i_{j-1}, o_j } (B^{i_{j-1}, o_j})^{\ast} \otimes B^{i_{j-1}, o_j}$~\cite{Orus2014,Schollwock2011}. $\overleftarrow{\transfer_j}$ denotes the action of $\transfer_j$ on a state from the left. 
Matrix multiplication is understood for the environment indices in Eq.(\ref{eq:rho}). Since the PPT in Eq.(\ref{eq:ppt}) is right-canonical, $\rho_j^E$ is related to $\Upsilon_{k:j+1, E}$ (the reduced density matrix of $\vert \Upsilon_{k:0}\rangle$ corresponding to sites $j+1$ to $k$, plus the environment $\alpha_k$) by an isometry (since each $B_{\alpha_{j-1}, \alpha_j}^{i_{j-1}, o_j}$ is an isometry from the Hilbert space $\mathcal{H}^E$ to $\mathcal{H}^E\otimes \mathcal{H}^S \otimes \mathcal{H}^S$), as a result $\rho_j^E$ has the same entanglement entropy as $\Upsilon_{k:j+1, E}$ (thus also the same as $\Upsilon_{j:0}$ since they are the two bipartition reduced density matrices from the PPT, which is a pure state). Therefore the memory complexity in Eq.(\ref{eq:renyi}) also measures the entropy of $\rho^E_j$. The importance of $\rho^E_j$ can be further seen by considering a `local' measurement (which means that we perform informationally complete preparations and measurements at all the previous time steps and then average over them) $\measure_{j}$ (written as a matrix $M_{o_j, o_j'}$)
\begin{align}
&\trace(M_j \ptensor_{j:0}) = \langle \Upsilon_{j:0} \vert M_j \vert \Upsilon_{j:0}\rangle \nonumber \\ 
=& \sum_{i_j, o_j, o_j', \alpha_{j-1}, \alpha_{j-1}', \alpha_j} \rho^E_{\alpha_{j-1}, \alpha_{j-1}'} B^{i_{j-1}, o_j}_{\alpha_{j-1}, \alpha_j} M_{o_j, o_j'} (B^{i_{j-1}, o_j'}_{\alpha_{j-1}', \alpha_j})^{\ast}  .
\end{align}
Therefore to measure a local observable at time step $j$, all one needs from the past is $\rho^E_{j-1}$. In other words, $\rho^E_{j-1}$ contains all the history information, and thus naturally corresponds to the distribution of the causal states in the $\epsilon$-machine. 

Now we can draw the connections between the classical stochastic process and the quantum process a step further. We have shown that the process tensor corresponds to the conditional probability on the histories, and the $\epsilon$-machine corresponds to the OQE model. From the discussions above, we can further see that the minimal environment corresponds to the space spanned by all the memory states in the $\epsilon$-machine, thus can be interpreted as the memory space. The memory size (the Schmidt rank of the PPT) is simply the size of the memory space. The effective environment state $\rho^E_j$ corresponds to the (stationary) distribution of the classical causal states.
The memory complexity corresponds to the classical memory complexity defined as the Renyi entropy of the (stationary) distribution of the causal states.


\begin{figure}
\includegraphics[width=\columnwidth]{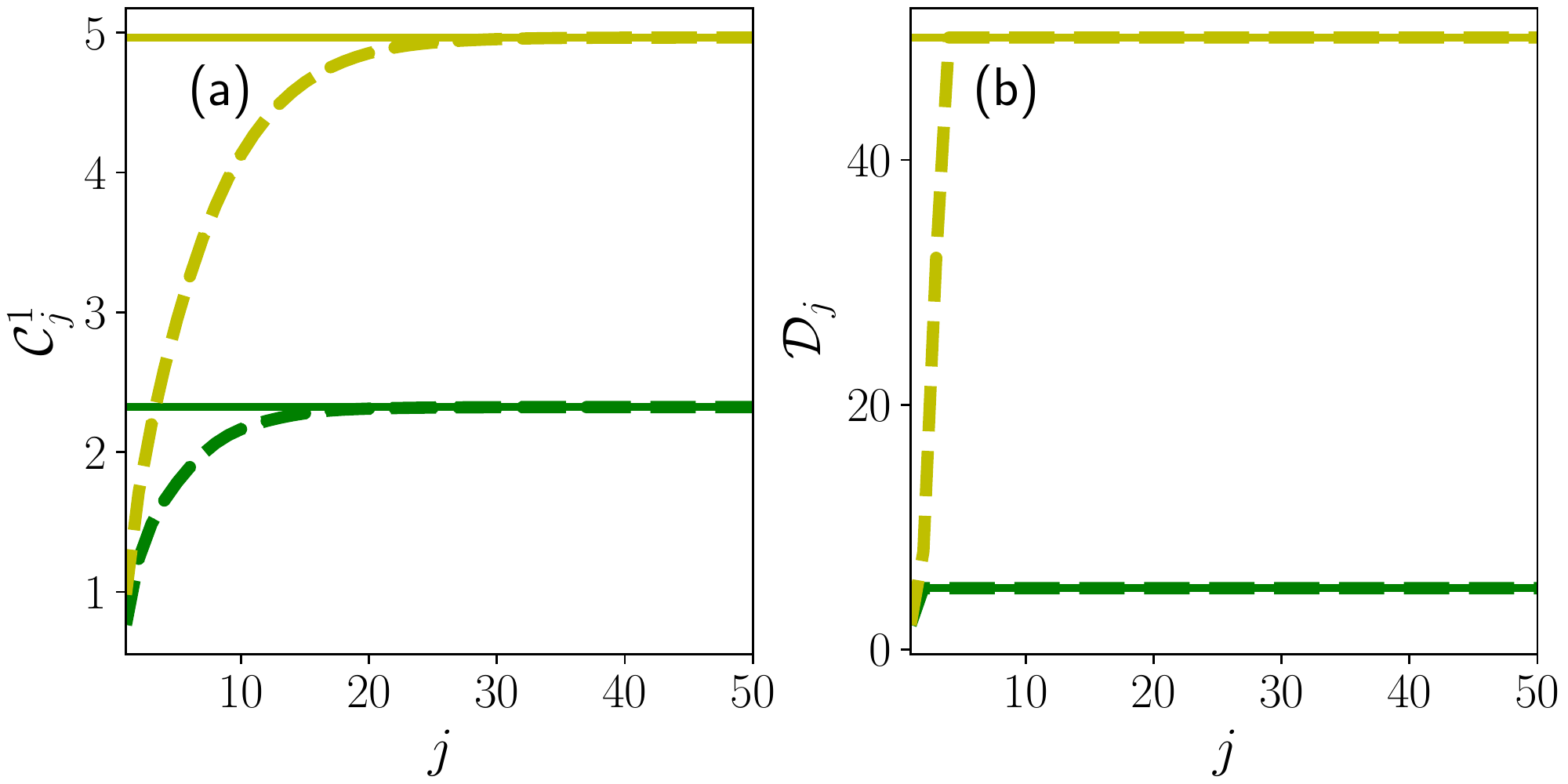}
\caption{(a) The memory complexity $\mem^{1}_{j}$ and (b) the memory size $\memsize_j$ as a function of the time step $j$. The yellow and green dashed lines are results for pure and mixed system-environment initial states respectively, while the yellow and green solid lines are the corresponding theoretical limits. We have used $d=2$, $D=5$. The unitary evolutionary operator for the hidden OQE model is randomly generated in the same way as Fig.~\ref{fig:fig3}. The pure and mixed system-environment initial states are also randomly generated.
}
\label{fig:fig4}
\end{figure}

Additionally, we have the following theorem for the memory complexity of a quantum process defined in Eq.(\ref{eq:ptdef}).

\textbf{Theorem 1.} Assuming that the non-Markovian open quantum dynamics is generated by a hidden OQE model which is time-independent with an environment size $D$, and that the dominate eigenstate of the transfer matrix $\transfer$ is non-degenerate, then $\memsize_{\infty}=D$, $\mem^{\gamma}_{\infty} = \log_2 (D)$ if the $\rho_0^{SE}$ is a pure state, and $\memsize_{\infty}=dD^2$, $ \mem^{\gamma}_{\infty} = \mathcal{C}_0^{\gamma} + \log_2 (D)$ if $\rho_0^{SE}$ is a mixed state with entanglement entropy $\mathcal{C}_0^{\gamma}$. 

\textit{Proof.} For pure $\rho_0^{SE}$, it suffices to show that $\rho^E_{\infty} =I^E/D$. For mixed $\rho_0^{SE}$ with purification denoted as $\vert \rho_0^{SE}\rangle = \sum_s \lambda_s \vert x_s \rangle \vert y_s\rangle $, where $\vert x_s\rangle$ is an external orthogonal basis set,  $\vert y_s\rangle$ an orthogonal basis set of the environment, and $\lambda_s$ the Schmidt numbers, it suffices to show that $\rho^{E'}_{\infty} = \sum_s \lambda_s^2 \vert x_s\rangle\langle x_s\vert \otimes I^E / D $, where $E'$ denotes the enlarged environment including the external basis. More details of the proof can be found in Appendix.~\ref{app:proof}. Interestingly, based on Theorem 1, one could use $\mem^{\gamma}_k$ to detect the memory size $\memsize_k$ for a quantum process with large enough $k$, since the former is experimentally accessible~\cite{DaleyZoller2012,IslamGreiner2015}.

From the above theorem we can see that the complexity of the process tensor tomography (thus also the complexity of the reconstruction algorithm for the hidden OQE model) shown in Sec.~\ref{sec:alg} are bounded by $O(kd^2 \memsize_{\infty}) = O(k d^2 2^{\mem^{\gamma}_{\infty}})$ (for time-independent $U$ we may be able to reconstruct the OQE model more efficiently with a very small $k$, as demonstrated in Fig.~\ref{fig:fig3}).
The growths of the memory complexity and the memory size with the time step are shown in Fig.~\ref{fig:fig4}, with the non-Markovian open quantum dynamics of the system generated by a hidden OQE model with a randomly generated unitary evolutionary operator. We can see that both of them converge to their limiting values predicted in Theorem 1.

\section{Conclusion}

In summary, we have presented an efficient algorithm for process tensor tomography, based on which one can reconstructed a minimal hidden OQE model for the non-Markovian open quantum dynamics of the system. We have defined the memory size and memory complexity for a quantum process, which characterize the complexity of the reconstruction algorithm, and which are closely related to the $\epsilon$-machine for classical stochastic process. Our algorithm can fully characterize generic non-Markovian quantum dynamics based only on experimentally measurable quantities, and it could be a useful technique to study non-Markovian noises on near-term quantum devices~\cite{RomeroHsieh2021}.


\begin{acknowledgments}
C. G. would like to thank Chengran Yang for helpful discussions. C. G. acknowledges support from National Natural Science Foundation of China under Grant No. 11805279.
\end{acknowledgments}

\bibliographystyle{apsrev4-1}

%

\appendix

\section{Detailed proof of Theorem 1}\label{app:proof}

In case $\evo$ is assumed time-independent, we have
\begin{align}
\rho^E_j =\overleftarrow{\transfer}^j(\rho^E_0),
\end{align}
as a result $\rho^E_j$ will converge to the left dominate eigenstate of $\transfer$ with the largest eigenvalue~\cite{Schollwock2011}. We assume that the dominate eigenstate of $\transfer$ is non-degenerate, which is similar to an ergodic requirement on $\evo$~\cite{YangGu2018}. 

First we will prove that any left eigenstate of $\transfer$ has an eigenvalue smaller or equal to $1$. 
We will also use a single index $\sigma_j$ to denote the tuple $(i_{j-1}, o_{j})$ for $j>0$ for briefness.
For any basis $\vert a\rangle \langle b \vert$ of the density matrix of the environment, we have
\begin{align}\label{eq:E1}
|\overleftarrow{\transfer}(\vert a\rangle\langle b\vert)|^2 = \sum_{\sigma, \sigma', \alpha, \alpha'} (B^{\sigma}_{a, \alpha'})^{\ast} B^{\sigma}_{b, \alpha} B^{\sigma'}_{a, \alpha'} (B^{\sigma'}_{b, \alpha})^{\ast}.
\end{align}
Now we define two matrices 
\begin{align}
X^{\sigma, \sigma'}_a &= \sum_{\alpha'} B^{\sigma}_{a, \alpha'} (B^{\sigma'}_{a, \alpha'})^{\ast} ; \\
Y^{\sigma, \sigma'}_b &= \sum_{\alpha} B^{\sigma}_{b, \alpha} (B^{\sigma'}_{b, \alpha})^{\ast},
\end{align}
which are semi-positive and Hermitian matrices by definition. We can also see that $\trace(X) = \sum_{\alpha', \sigma} B^{\sigma}_{a, \alpha'} (B^{\sigma}_{a, \alpha'})^{\ast} = 1 $ since  $B$ is right-canonical, and the same for $Y$.
Then Eq.(\ref{eq:E1}) can be written as
\begin{align}\label{eq:proof1}
|\overleftarrow{\transfer}(\vert a\rangle\langle b\vert)|^2 &= \trace(X^{\dagger} Y) \leq \sqrt{\trace(X^{\dagger} X)}\sqrt{\trace(Y^{\dagger} Y)} \nonumber \\ 
&= \sqrt{\trace(X^2)} \sqrt{\trace(Y^2)} \nonumber \\ 
&\leq \sqrt{\trace^2(X)} \sqrt{\trace^2(Y)} \nonumber \\ 
&= \trace(X) \trace(Y) = 1 ,
\end{align}
where the second step in the first line of Eq.(\ref{eq:proof1}) follows from the Cauchy-Schwarz inequality and the inequality in the second line is due to the semi-positivity of $X$ and $Y$. Equality holds only if $a = b$. Thus for any state $\rho = \sum_{a, b}\rho_{a, b}\vert a\rangle\langle b\vert$, we have
\begin{align}
|\overleftarrow{E}(\rho)|^2 = |\sum_{a, b} \rho_{a, b} \overleftarrow{E}(\vert a\rangle\langle b\vert)|^2 \leq |\rho|^2 |\overleftarrow{E}(\vert a\rangle\langle b\vert)|^2  = |\rho|^2, 
\end{align}
therefore any left eigenvector of $\transfer$ has an eigenvalue that is not greater than $1$.

Second we show that the maximally mixed state $\hat{I}^E / D$ is both a left and right eigenvector of $\transfer$ with eigenvalue $1$. This immediately follows since $B^{\sigma}$ is both left and right-canonical (except for the first site, $B^{o_0}$, which does not matter for large time steps). Thus the first part of Theorem 1 is proved. 

Now we proceed to prove the second part of Theorem 1 for mixed system-environment initial state. In this case we need to first purify the initial state with $dD$ external basis $\vert x_s\rangle$ as 
\begin{align}\label{eq:purify}
\vert \rho^{SE}_0\rangle = \sum_s \lambda_s \vert x_s\rangle \vert y_s\rangle.
\end{align}
Accordingly the site matrix should be enlarged to
\begin{align}
B^{\sigma'} = B^{\sigma} \otimes I^{SE}.
\end{align}
The enlarged transfer matrix $\transfer' = \sum_{\sigma'} (B^{\sigma'})^{\ast} \otimes B^{\sigma'} $ is certainly degenerate. 
We note that the traceless matrices span a linear subspace which is orthogonal to the maximally entangled state and that the transfer matrix only maps traceless matrices to traceless matrices due to trace preservation, then since the largest eigenvalue is assumed to be non-degenerate, all the traceless matrices must have eigenvalues strictly less than $1$. 
From Eq.(\ref{eq:purify}), the enlarged effective environment state $\rho^{E'}_0$, which includes the original environment plus the external basis $|x_s\rangle$, can be written as
\begin{align}
\rho^{E'}_0 =  \sum_{s, s'} \lambda_s \lambda_{s'} \vert x_s\rangle\langle x_{s'}\vert \otimes \trace_S\left(\vert y_s\rangle  \langle y_{s'}\vert\right).
\end{align}
Then we have
\begin{align}\label{eq:rhoEj}
\rho^{E'}_j &=  \overleftarrow{T'}^j(\rho^{E'}_0) \nonumber \\ 
&= \sum_{s, s'} \lambda_s \lambda_{s'} \vert x_s\rangle\langle x_{s'}\vert \otimes \overleftarrow{\transfer'}^j( \trace_S(\vert y_s\rangle  \langle y_{s'}\vert)).
\end{align}
For $s\neq s'$, we have $\trace_E(\trace_S(\vert y_s\rangle  \langle y_{s'}\vert)) = \langle y_{s'}\vert y_s\rangle = 0$, namely $\trace_S(\vert y_s\rangle  \langle y_{s'}\vert)$ is a traceless density matrix of the environment. Thus from the previous arguments the state $\trace_S(\vert y_s\rangle  \langle y_{s'}\vert)$ lives in a subspace with eigenvalue strictly less than $1$, and we have $\overleftarrow{\transfer'}^{\infty}( \trace_S(\vert y_s\rangle  \langle y_{s'}\vert)) = 0$. As a result, for $j\rightarrow \infty$, Eq.(\ref{eq:rhoEj}) becomes
\begin{align}
\rho^{E'}_{\infty} = \sum_s \lambda_s^2 \vert x_s\rangle\langle x_s\vert \otimes I^E / D.
\end{align}
Thus the second part of Theorem 1 has been proved.

\end{document}